\documentclass[10pt,aps,prd,nobibnotes,twocolumn,
nobalancelastpage,superscriptaddress]{revtex4}
\usepackage{graphicx}
\usepackage{amsmath}
\def\gap{\;\rlap{\lower 2.5pt
 \hbox{$\sim$}}\raise 1.5pt\hbox{$>$}\;}
\def\lap{\;\rlap{\lower 2.5pt
   \hbox{$\sim$}}\raise 1.5pt\hbox{$<$}\;}
\def\gsim{\;\rlap{\lower 2.5pt
 \hbox{$\sim$}}\raise 1.5pt\hbox{$>$}\;}
\def\lsim{\;\rlap{\lower 2.5pt
   \hbox{$\sim$}}\raise 1.5pt\hbox{$<$}\;}

\newcommand{\crosssec}{\sigma_{\tilde\chi^0_1-p}^{\rm SI}}
\newcommand{\crosssecsd}{\sigma_{\tilde\chi^0_1-p}^{\rm SD}}

\def\bone{B^{(1)}}

\def\lsim{\raise0.3ex\hbox{$\;<$\kern-0.75em\raise-1.1ex\hbox{$\sim\;$}}}
\def\gsim{\raise0.3ex\hbox{$\;>$\kern-0.75em\raise-1.1ex\hbox{$\sim\;$}}}

\begin{document}


\title{Dark Matter: A Multidisciplinary Approach}

\author{Gianfranco Bertone}
\affiliation{Institut d'Astrophysique de Paris, UMR 7095-CNRS,
Universit\'{e} Pierre et Marie
Curie, 98bis boulevard Arago, 75014 Paris, France}
\email{bertone@iap.fr}
\pacs{}
\date{\today}

\begin{abstract}
We review the current status of accelerator, direct and
indirect Dark Matter (DM) searches, focusing on the complementarity
of different techniques and on the prospects for discovery. 
After taking a census of present and upcoming DM-related 
experiments, we review the motivations
to go beyond an "accelerator-only" approach, 
and highlight the benefits of multidisciplinarity in the 
quest for DM.
\end{abstract}

\maketitle


\section{Introduction}

The evidence for non-baryonic dark matter is compelling at all 
observed astrophysical scales~\cite{Bergstrom:2000pn,Bertone:2004pz}. Although alternative explanations 
in terms of modified gravity (see Ref.~\cite{Bekenstein:2004ne} for a relativistic theory of the MOND paradigm) cannot be ruled out, they can hardly
be reconciled with the most recent astrophysical observations~\cite{Clowe:2006eq} without requiring additional matter beyond the observed 
baryons (e.g. Ref.~\cite{Feix:2007zm} and references therein).
It is therefore natural to ask {\it how can we identify the nature of 
DM particles?}. We review here the 
main strategies that have been devised to attack this problem, namely
accelerator, direct and indirect searches, focusing on the interplay
between them and on their complementarity.

In fact, a tremendous theoretical and experimental effort is in progress to 
clarify the nature of DM, mostly devoted, but not limited, to 
searches for 
Weakly Interacting Massive Particles (WIMPs), 
that achieve the appropriate relic density by {\it freezing-out} of 
thermal equilibrium when their self-annihilation rate becomes smaller 
than the expansion rate of the Universe. 
The characteristic mass of these particles is $\cal{O}$$(100)$ GeV, and
the most representative and commonly discussed candidates in this class 
of models are the supersymmetric neutralino, and the B$^{(1)}$ particle, 
first excitation of the hypercharge gauge boson, in theories with 
Universal Extra Dimensions.

A tentative census of present and upcoming DM experiments (WIMPs only) 
is shown in fig. 1. Shown in the figure are:  
two particle accelerators, viz. the Tevatron at Fermilab, and the upcoming
Large Hadron Collider (LHC) at CERN; the many direct detection experiments
currently taking data or planned for the near future, along with the names
of the underground laboratories hosting them; high-energy neutrino 
telescopes; gamma-ray observatories; gamma-ray and anti-matter satellites.
Light blue points denote gamma-ray experiments that are not 
directly related to indirect DM searches, as DM signals would be typically
produced at energies below their energy threshold. Nevertheless, they may 
turn out to be useful to discriminate the nature of future 
unidentified high-energy gamma-ray sources.
Three satellites are shown in the inset of figure 1: PAMELA, an anti-matter
satellite that has already been launched and is expected to release the first
scientific data very soon. ; GLAST, a gamma-ray satellite that is scheduled for launch in early 2008; and AMS-02, anti-matter satellite that should 
be launched in the near future. 

We will discuss below the prospects for detecting
DM with the various experiments shown in fig. 1, and we will focus our attention 
on the complementarity of the various detection strategies. The paper is 
organized as follows: we first discuss accelerator searches, and show that although the LHC has
the potential to make discoveries of paramount importance for our understanding
of DM, it may not be able to solve all problems. In Section 3 we discuss
the information that can be extracted from direct detection experiments, 
in case of positive detection. Section 4 is then dedicated to indirect searches, 
and to the question of what astrophysical observations can tell
us about the nature of DM, and how to combine this information with all other
searches.

\begin{figure*}
\centering
  \includegraphics[width=\textwidth]{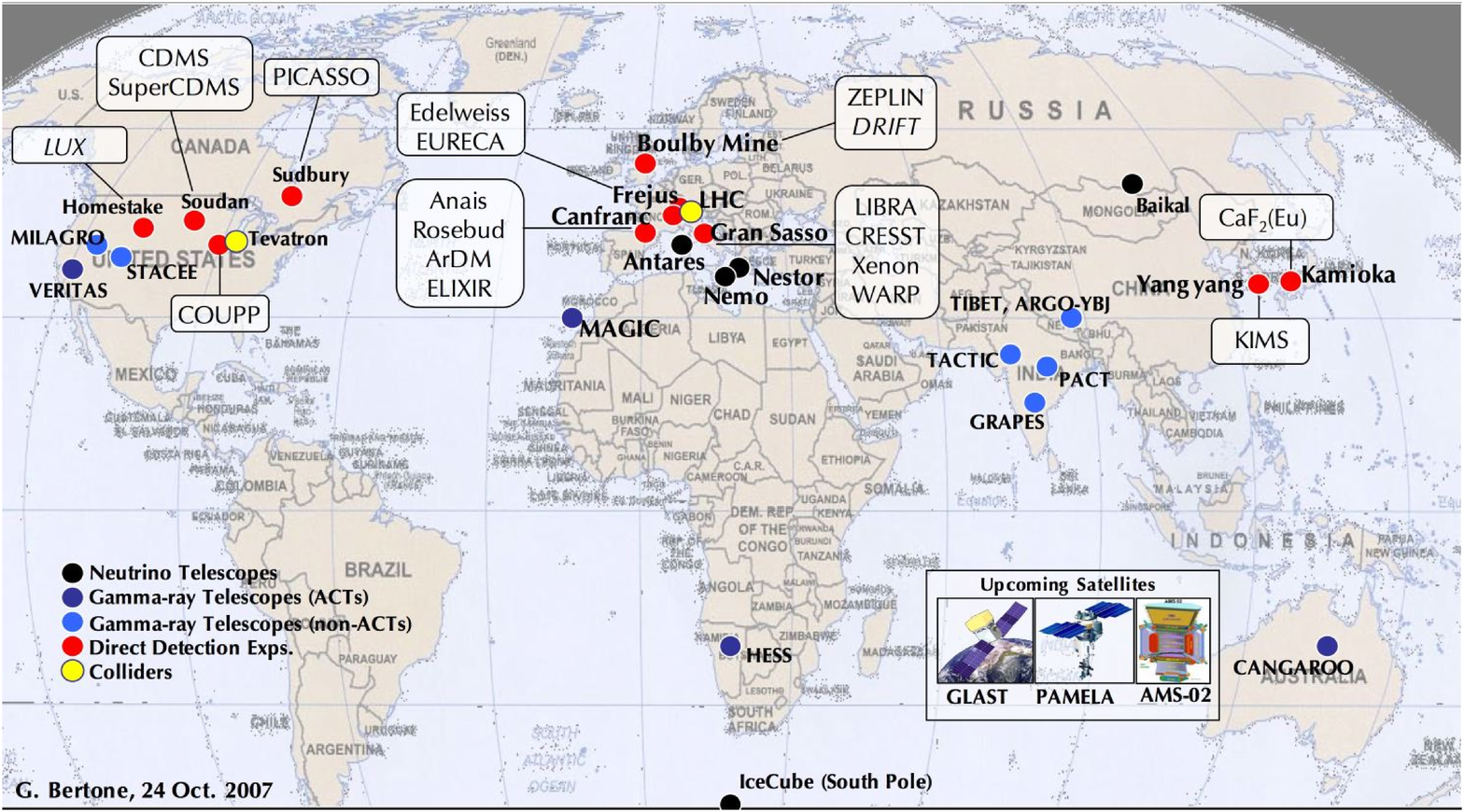}
\caption{2007 census of present and upcoming Dark Matter-related 
experiments. Black points 
denote the location of high energy neutrino telescopes; Dark-blue points 
are for gamma-ray Air Cherenkov Telescopes, while light-blue points
are for other ground-based gamma-ray observatories. Red points are 
for underground laboratories hosting existing and upcoming direct 
detection experiments. Yellow points show the location of the 
Fermilab's Tevatron, and the upcoming Large Hadron Collider at CERN.}
\label{DMmap}       
\end{figure*}

\section{Colliders may not be enough}

The Large Hadron Collider (LHC), which is about to start 
operations at CERN, will allow us to explore possible 
extensions of the Standard Model of particle physics,
reaching a center-of-mass energy of about 14 TeV. Obviously,
the discovery of new particles would be of paramount 
importance also for Astrophysics and Cosmology, and would
represent a big step forward in our understanding of
the Universe. However, as we argue below, it will not 
be easy to extract, from 
accelerator experiments alone, enough information to 
unambiguously identify DM particles.  
 
The constraints that can be placed on a dark matter 
candidate from collider experiments are strongly model-dependent, and 
it is, unfortunately, impossible to describe the reach of 
colliders in their search for dark matter in any kind of general way. 
However, a number of searches for particles associated with a dark 
matter candidate already provide interesting constraints on
proposed extensions of the Standard Model of particle physics, and
they include studies such as: invisible $Z$ width, Sneutrino limits,
searches for new charged or colored particles, searches
for the Higgs or new gauge bosons, flavor changing neutral currents, 
$b \rightarrow s \gamma$, $B_s \rightarrow \mu^+ \mu^-$, 
anomalous magnetic moment of the muon and electroweak precision 
measurements.

Together, these constraints can be very powerful, often providing very tight bounds for specific models. 
The LHC will  test numerous classes of models, searching at scales of up to several TeV. In addition to the Higgs boson(s), the LHC will be in particular sensitive to most supersymmetry scenarios, to models with TeV-scale universal extra dimensional and to little Higgs models, which are three examples of classes of models with "natural" Dark Matter candidates. (see e.g. Refs.\cite{atlas,cms,Dawson:1983fw,Allanach:2004ub,Allanach2,lhcreach1,lhcreach4,lhcreach6,lhcreach7}). 

\begin{figure}[t]
\centering
  \includegraphics[width=0.2\textwidth]{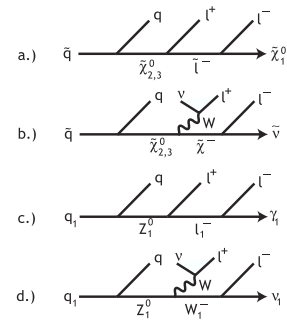}
\caption{Four scenarios for decay chains observed at LHC.  Each exhibits jets,
hard leptons, and missing energy.  Distinguishing between these cases may not be possible with LHC alone. From Ref.~\cite{Baltz:2006fm}}
\label{fig:diagrams}       
\end{figure}

\begin{figure*}[t]
\centering
\includegraphics[width=0.51\textwidth]{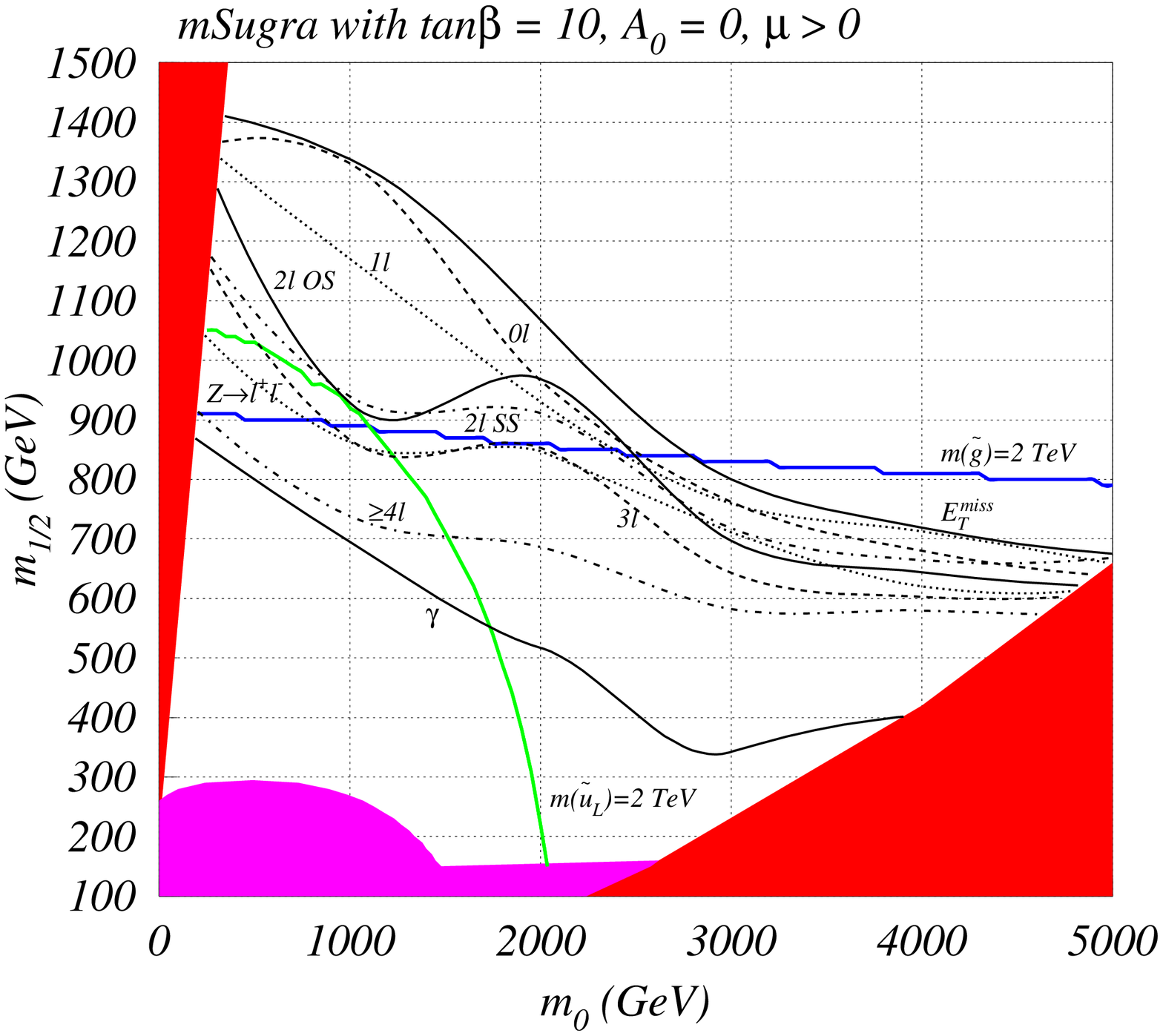}
  \includegraphics[width=0.47\textwidth]{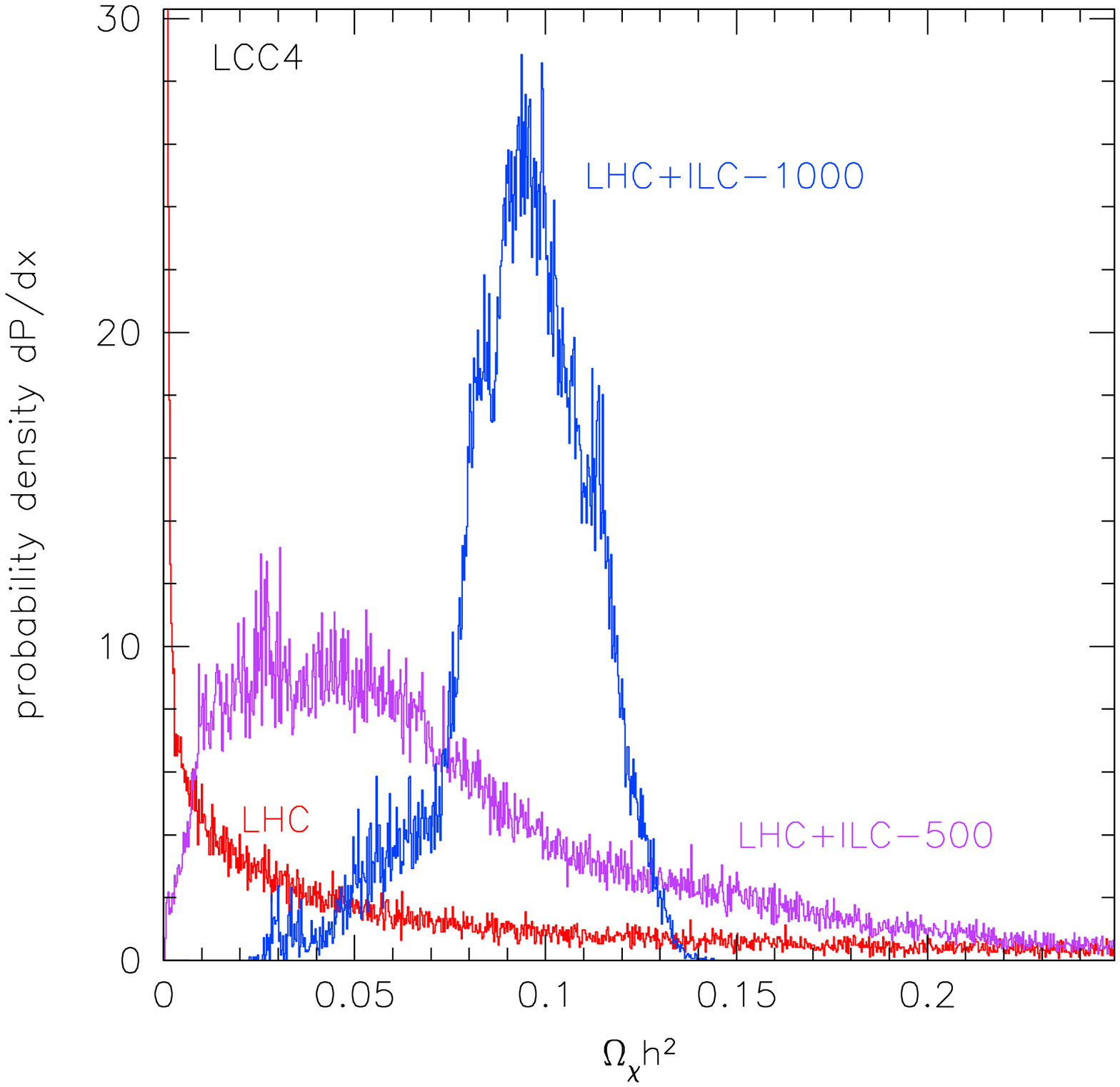}
\caption{{\it Left Panel.} The reach of the Large Hadron Collider (LHC) in the $(m_0,m_{1/2})$ plane of the mSUGRA scenario, with $\tan \beta = 10$, $A_0=0$ and positive $\mu$, for a variety of channels. Also shown are the 2 TeV up squark and 2 TeV gluino mass contours. The red regions are excluded by theoretical constraints, while the magenta region is excluded experimentally. $100 \rm{fb}^{-1}$ of integrated luminosity is assumed. From Ref.~\cite{lhcreach1}. {\it Right Panel.} Relic density measurement for one of the four benchmark points discussed in \cite{Baltz:2006fm}. Histograms give the probability distribution  for the
reconstructed $\Omega h^2$, given three different sets of accelerator constraints. 
Results for the LHC make use of the assumption that the underlying physics model
is supersymmetry (which might not emerge clearly from the LHC data
alone). The difficulty in reconstructing the relic density of a WIMP,
based on LHC data only is apparent, suggesting the need of a multidisciplinary
approach to DM searches. From Ref. \cite{Baltz:2006fm}}
\label{fig:collider2}     
\end{figure*}

In order to show the potential and the inevitable limits of an 
accelerator-only approach, we focus on Supersymmetry (SUSY),
which is the most studied, and possibly the most promising, 
extension of the Standard Model of particle physics.
In SUSY, a discrete symmetry, known as $R$-parity,
is often imposed in order to forbid lepton and baryon violating
processes which could lead, for instance, to proton decay. As a 
consequence, SUSY particles
are only produced or destroyed in pairs, thus making the lightest
supersymmetric particle (LSP) stable. Remarkably, in large areas of
the parameter space of SUSY models, the LSP is an electrically neutral
particle, the lightest neutralino, $\tilde{\chi}^0_1$, which therefore
constitutes a very well motivated candidate for dark matter, within
the class of WIMPs. 
In the left panel of Fig.~\ref{fig:collider2}, the reach of the LHC is shown \cite{lhcreach1}. It is interesting to note that in the region of the MSSM which is the most difficult to probe at the LHC, direct dark matter detection rates are very high \cite{directwherenolhc}.

Since we cannot observe with the LHC the final-state
WIMPs, we cannot learn the energies and momenta of the produced 
particles from the final state. Without knowing the rest frame of the 
massive particles, it is then very difficult to determine the spins of these
particles or to specifically identify their decay modes. 
In Fig.~\ref{fig:diagrams} (taken from Ref. \cite{Baltz:2006fm}) we 
show four models of 
the decay of a colored primary particle. 
Examples (a) and (b) are drawn from models of supersymmetry in which the 
WIMP is the supersymmetric partner of the photon or neutrino.  Examples
(c) and (d) are drawn from models of extra dimensions in which the 
WIMP is, similarly, a higher-dimensional excitation of a photon or a 
neutrino. 
The observed particles in all four decays are the same and the 
uncertainty in reconstructing the frame of the primary colored particle
make it difficult to discriminate the 
subtle differences in their momentum distributions.
Although model-dependent features can help distinguishing the cases 
of supersymmetry and
extra dimensions \cite{Datta:2005zs,Smillie:2005ar} it is unlikely that the
LHC will unambiguously identify the nature of the WIMPs.

Even if new particles are identified, it might be difficult to 
understand whether they can account for {\it all} the DM in the Universe.
In fact, it will be necessary to infer their relic density from 
physical quantities measured at accelerator, a process that will
require some assumptions on the particle physics and 
cosmological setup. Even assuming, say, a minimal supersymmetric
scenario and a standard expansion history of the Universe, this
program may turn out to be impossible with LHC data only. 

A detailed study, along these lines, has been performed by Baltz
et al.~\cite{Baltz:2006fm}, who have analyzed 4 benchmark points in 
a minimal supersymmetric scenario, showing that only a poor 
reconstruction of the relic density will be possible in most
cases. We show in the right panel of fig.~\ref{fig:collider2}
the probability distribution for the DM relic density,
for their benchmark point LCC4, which is chosen in a region 
where the $A^0$ resonance makes an important contribution to 
the neutralino annihilation cross section. In the same figure, 
the improvements in the determination of the relic densities 
with two different versions of a future Linear Collider are 
also shown. 

This case is representative of a large portion of the
theory parameter space where the reconstruction of the relic
density of DM particles cannot be performed at the level of
accuracy that matches existing and upcoming CMB experiments
such as WMAP and Planck. 

Furthermore, one has to keep in mind that
unexpected surprises may await us, such as a non Standard expansions
history of the Universe (in which case our relic density calculations
should be revised) or a much more complicated "dark sector" in 
extensions of the Standard Model of particle Physics (in which
case our calculations make no sense at all). 

It is therefore 
crucial to perform all possible searches, including direct and
indirect DM searches discussed below, and to devise 
strategies that would allow us to combine the information from
a large set of diverse experiments into a consistent theoretical 
scenario.

\section{Getting the most out of Direct Searches}
\begin{figure}[t]
\centering
  \includegraphics[width=0.45\textwidth]{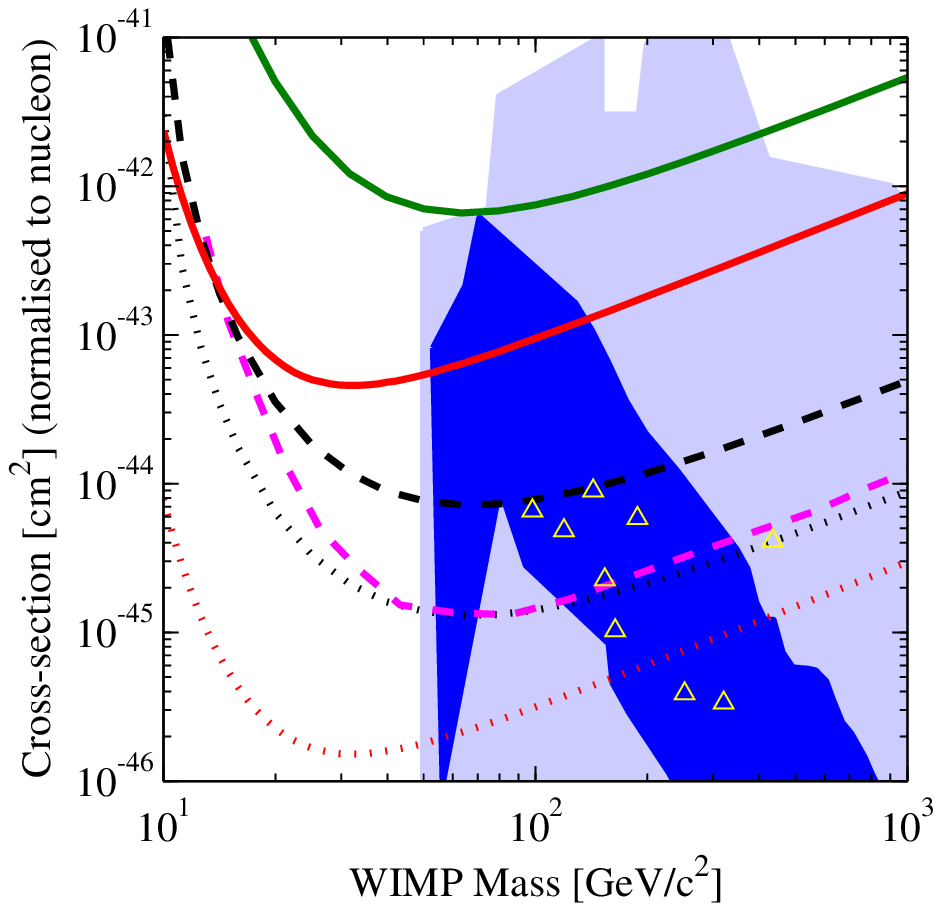}
\includegraphics[width=0.5\textwidth]{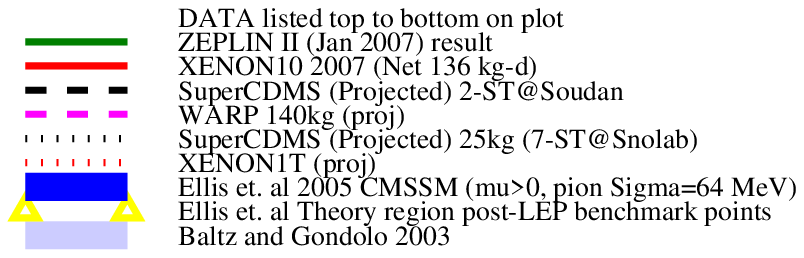}
\caption{Sensitivity of current and planned direct 
detection experiments~\cite{dmtools}.}
\label{fig:direct}       
\end{figure}

DM can be
searched for {\it directly}, as DM particles passing through the
Earth interact inside large detectors. The field of direct searches is
well-established, with many experiments currently operating 
or planned (see fig.\ref{fig:direct} for an example of the reach of present 
and upcoming direct detection experiments in the $\sigma_{\chi N}$--$m_\chi$ plane). 
\begin{figure*}[t]
  \includegraphics[width=0.49\textwidth]{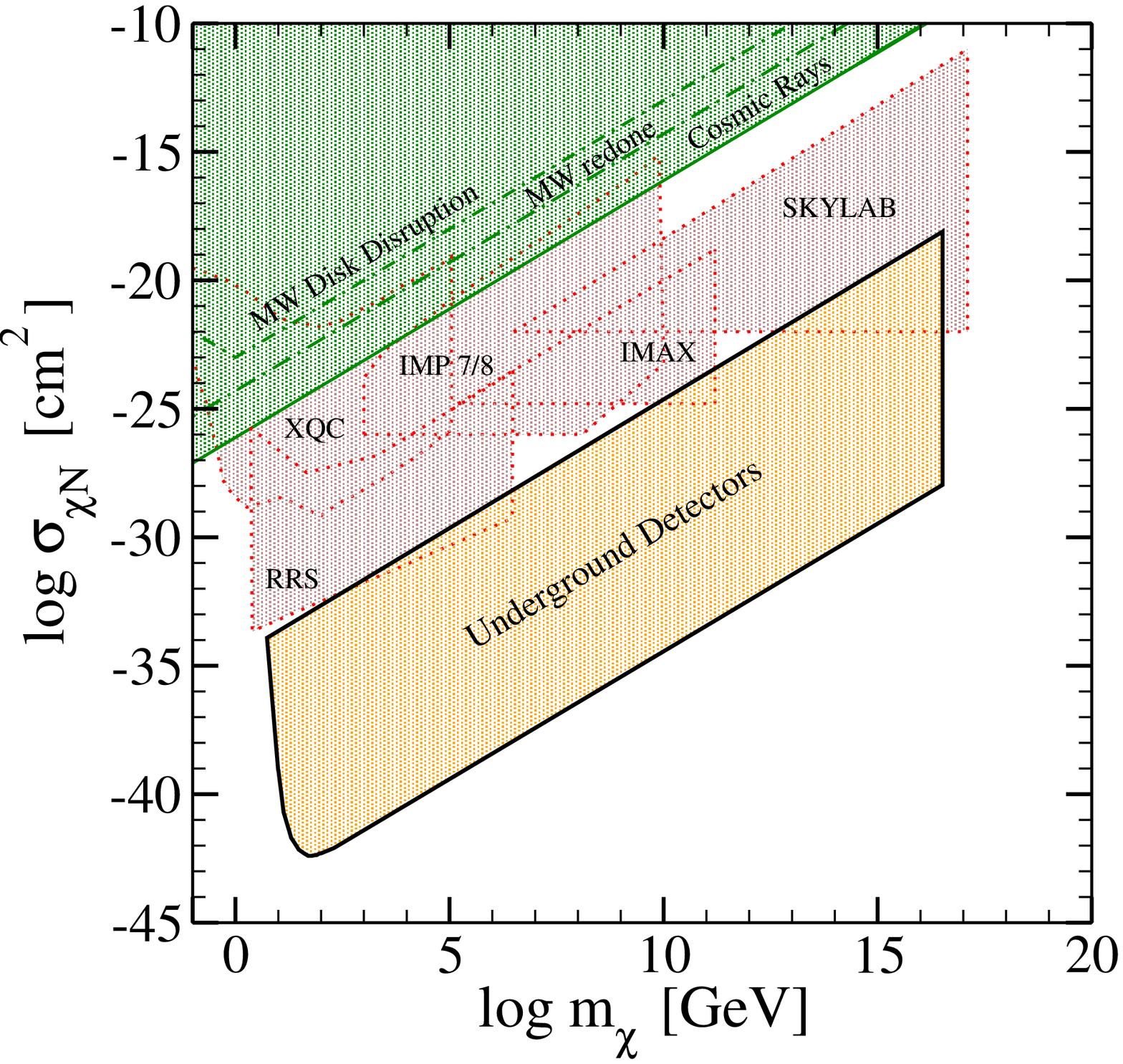}
  \includegraphics[width=0.49\textwidth]{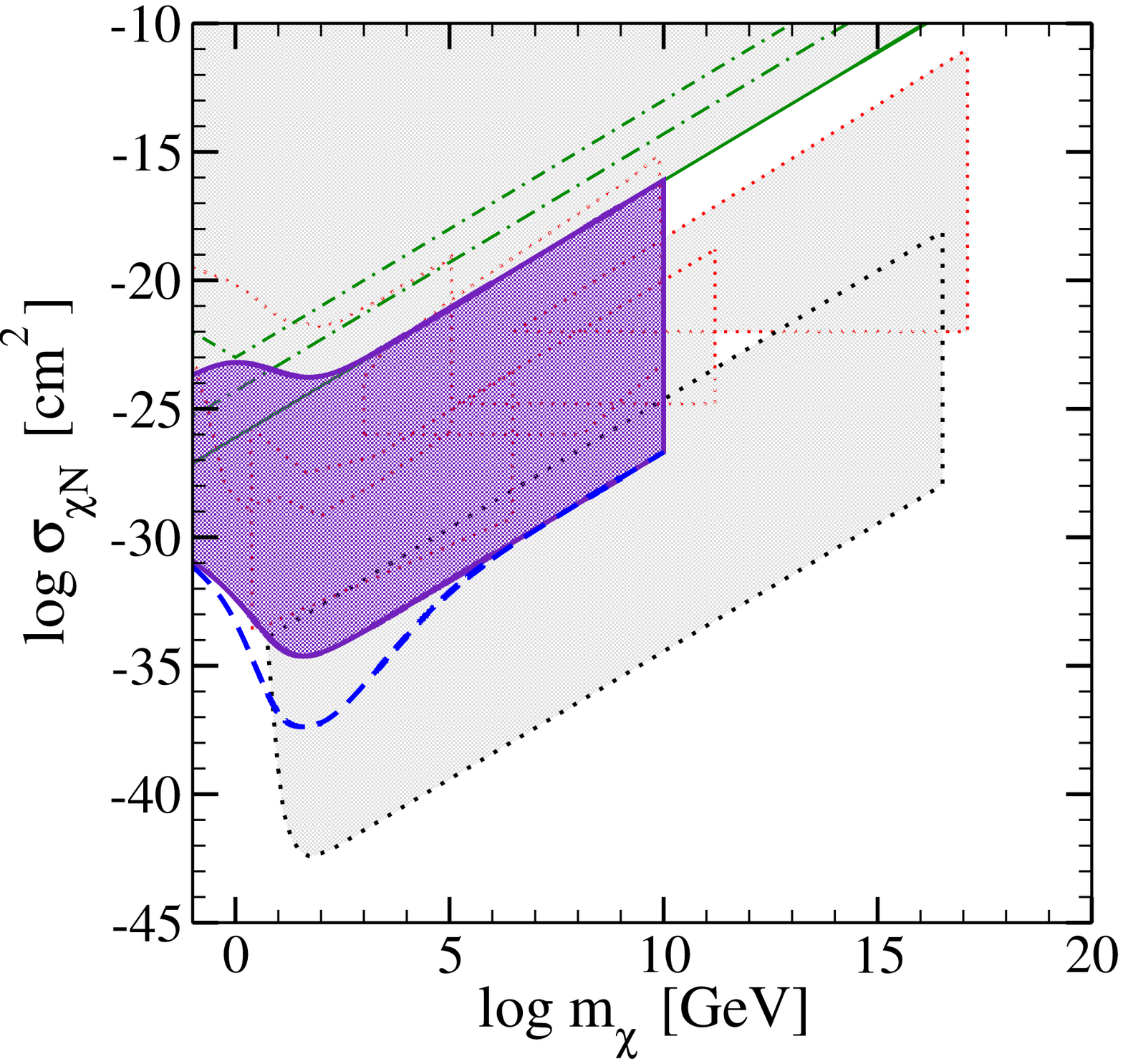}
  \caption{{\it Left Panel.} Excluded regions in the
$\sigma_{\chi N}$--$m_\chi$ plane.  From top to bottom, these come from astrophysical
constraints
(dark-shaded),
re-analyses of high-altitude detectors
(medium-shaded),
and underground direct dark matter detectors
(light-shaded). 
{\it Right Panel.} Inside the heavily-shaded region, dark
  matter annihilations would overheat Earth.  Below the top edge of
  this region, dark matter can drift to Earth's core in a satisfactory
  time.  Above the bottom edge, the capture rate in Earth is nearly
  fully efficient, leading to a heating rate of 3260 TW (above the
  dashed line, capture is only efficient enough to lead to a heating
  rate of $\gtrsim$ 20 TW). Both figs. from Ref.~\cite{Mack:2007xj}.
}
  \label{earth}
\end{figure*}

The idea is to measure the recoil energy of nuclei 
hit by DM particles in large detectors. The rate at which
these interactions occur is typically very small, and it 
is approximately given by 
\begin{equation}
R\approx \sum_i N_i n_\chi <\sigma_{i \chi}>,
\end{equation}
where the index, $i$, runs over nuclei species present in
the detector,  $N_i$ is the number of target nuclei in the detector,
$n_\chi$
is the local WIMP density and $<\sigma_{i \chi}>$ 
is the cross section for the scattering of WIMPs off 
nuclei of species $i$, averaged over the relative 
WIMP velocity with respect to the detector (see e.g. 
Ref.~\cite{Jungman:1995df}). 

Direct detection relies on scalar, or {\it spin-independent}, couplings,
describing coherent interactions of DM with the entire nuclear mass, and
on axial, or {\it spin-dependent}, couplings, describing the
interaction of DM with the spin-content of the nucleus. 
Experimental
efforts have so far focused on targets which enhance
the scalar-interaction scattering rate, but as we shall see below, 
the quantification of both types of interactions, 
by measurement of the scattering cross-section on multiple target nuclei
significantly improves our ability to identify the nature of DM particles.
We recall here that spin-dependent couplings are also important
for indirect DM searches, as detection of high energy neutrinos from 
DM annihilations at the center of the Sun, is only possible for candidates
with large enough spin-dependent interactions with the nuclei of the Sun 
(see below for further details, and see also Ref.~\cite{HooperTaylor}).

Before discussing the details of direct detection, we recall that
if DM exists in the form of particles, its interactions must be 
truly weak, for a very wide range of masses.  In fact, a new and 
largely model-independent constraint on the dark matter scattering 
cross section with nucleons, which actually relies on 
an indirect-detection approach, was recently derived in Ref.~\cite{Mack:2007xj}. When the dark matter capture 
rate in Earth is efficient, the rate of energy deposition by dark 
matter self-annihilation products would grossly exceed the measured 
heat flow of Earth. This improves the spin-independent cross section 
constraints by many orders of magnitude, and closes the window between 
astrophysical constraints (at very large cross sections) and underground 
detector constraints (at small cross sections). For DM masses between 
1 and $10^{10}$ GeV, the scattering cross section 
of dark matter with nucleons is then bounded from above by the latter 
constraints, and hence must be truly weak, as usually assumed. 
In fig.~\ref{earth} we show the regions of the $\sigma_{\chi N}$--$m_\chi$ plane
where dark matter annihilations would overheat Earth, along with the 
astrophysical and ground-based constraints (see Ref.~\cite{Mack:2007xj}
for further details).

A natural question to ask is to what accuracy direct detection
experiments can determine parameters such as the DM mass and the
scattering cross-section off nucleons. Furthermore, it is 
important to understand how these parameters can be used to 
identify the underlying theoretical framework, e.g. discriminating
neutralino DM from other candidates. 
It was recently shown  that assuming a DM particle with scattering 
cross section of $10^{−7}$ pb, i.e. just below current exclusion 
limits, and fixing the local DM velocity distribution and density,
an exposure of $3 \times 10^3$ ($3 \times 10^4, 3 \times 10^5$) kg day
is needed in order to measure the
mass of a light DM particle with an accuracy of roughly 25\% (15\%,2.5\%)~\cite{Green:2007rb}.
This corresponds more or less to the three proposed phases of SuperCDMS.
These numbers increase with increasing WIMP mass, and for DM particles 
heavier than 500GeV, even with a large exposure it will only be possible 
to place a lower limit on the mass~\cite{Green:2007rb}. 

Recently, the need of combining spin-dependent and -independent 
techniques in order to effectively identify the nature of DM,
has been emphasized in Ref.~\cite{Bertone:2007xj}.

\begin{figure*}[t]
 \center
  \includegraphics[width=0.98\textwidth]{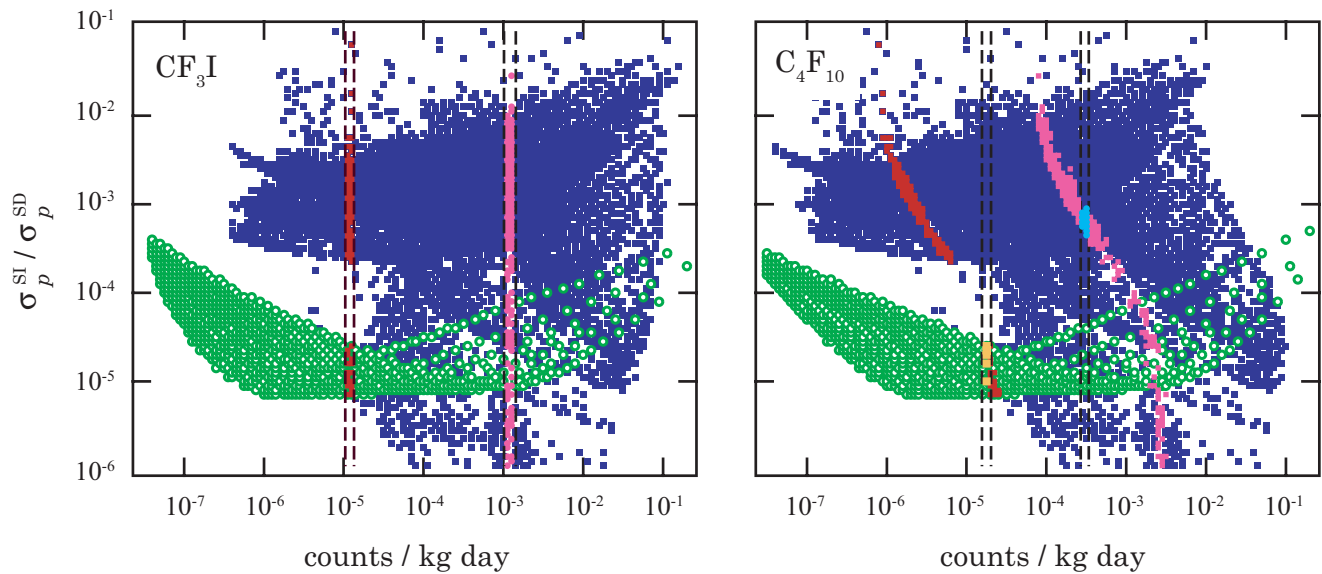}
  \caption{ 
 {\it Left Panel:} The detection of a DM signal with a CF$_3$I
    detector can only loosely constrain DM candidates (blue squares
    for neutralinos, green circles for the LKP) in the
    $\sigma^{\rm SI}_p / \sigma^{\rm SD}_p$ versus count-rate
    plane. Red (magenta) dots show the many models consistent with a
    measurement of  $\sim10^{-5}$ ($10^{-3}$) counts / kg day on
    CF$_3$I. {\it Right Panel:} measurement of the event rate in a
    second detection fluid such as C$_4$F$_{10}$, with lower
    sensitivity to spin-independent couplings, effectively reduces the
    remaining number of allowed models--orange (aqua) dots--and
    generally allows discrimination between the neutralino and the
    LKP (a 10\% uncertainty in the measurements is adopted here for
    illustration). From Ref.~\cite{Bertone:2007xj}}
  \label{targets}
\end{figure*}

To be more precise, let us focus on two specific classes of DM candidates.
The first one is the aforementioned neutralino, arising in supersymmetric
extensions of the Standard Model of particle physics. 
In order to determine the theoretical predictions for the neutralino 
detection cross section we follow the analysis in Ref.~\cite{Bertone:2007xj}
where a random scan in the effective MSSM  (effMSSM)
scenario, with input quantities defined at the electroweak
scale
\cite{effMSSM} has been performed. The mass parameters
have been taken in the range $0\le\mu$, $m_A$, $M_1$, $A$,
$m\le 2$~TeV with $3\le\tan\beta\le50$.
A small non-universality in squark soft masses has also been included,
taking $m^2_{Q,u,d}=(1\ldots5)\,m^2$. Noteworthy,
regions with large $\crosssecsd$ are obtained, some of
which predict a small $\crosssec$. 
A second scan was performed in the framework of supergravity-inspired
models in which the soft terms are inputs at the grand
unification scale. We consider the most general situation,
with non-universal scalar and gaugino masses,
exploring the scenarios presented in
\cite{nunivsugra} for $3\le\tan\beta\le50$.
Another theoretically very well motivated candidate arises in
theories  with Universal Extra Dimensions (UED), in which all 
fields are allowed to propagate in the bulk \cite{Appelquist:2000nn}.
In this case, the Lightest Kaluza-Klein Particle (LKP) is a
viable DM candidate, likely to be associated with the first KK excitation 
of the hypercharge gauge boson~\cite{Cheng:2002iz,Servant:2002aq}, 
usually referred to as $\bone$. 
In absence of spectral degeneracies, the $\bone$ would achieve the 
appropriate relic density for masses in the 850--900 GeV range~\cite{Servant:2002aq}. 
Interestingly, due to the quasi-degenerate nature of the KK spectrum, this
range can be significantly
modified, due to coannihilations with first~\cite{Burnell:2005hm,Kong:2005hn}
and second~\cite{Kakizaki:2005en,Matsumoto:2005uh,Kakizaki:2006dz} KK-level 
modes. The allowed mass range was also found to depend significantly on 
the mass of the Standard Model Higgs boson~\cite{Kakizaki:2006dz}, and 
in general on the matching contributions to the brane-localized kinetic 
terms at the cut-off scale (see the discussion in Ref.~\cite{Burnell:2005hm}). 
The LKP models tend to populate a different region of 
the parameter space with respect to SUSY scenarios, because of the
larger spin-dependent cross-section.

Supposing an experiment succeeds in directly detecting DM particles,
it is interesting to consider how the nature of the DM (e.g.
neutralino or LKP) might be determined.  The possibility of 
combining the information from different detection targets 
make it possible 
to determine the nature of DM, upon successful detection, with much
better accuracy ~\cite{Bertone:2007xj}. As shown in
Fig.~\ref{targets}(a), in fact, the measurement of an event rate in 
an experiment such as the Chicagoland Underground Observatory (COUPP)
~\cite{COUPP},  does reduce allowed models, but does not generally
place significant constraints on coupling parameters or on the nature
of detected DM (i.e. neutralino or LKP). 

However, as shown in
Fig.\,\ref{targets}b), subsequent detection of an event rate on a
second target does substantially reduce the allowed range of
coupling parameters, and allows, in most cases, an effective discrimination
between
neutralino and LKP dark matter. The combination of detector fluids
used in Fig.\,\ref{targets} is effective in reducing the allowed
range of $\sigma^{\rm SI}_p / \sigma^{\rm SD}_p$ because massive
iodine nuclei have a large SI coupling, while fluorine nuclei have a
large SD$_p$ coupling. It must be noted that fluorine and iodine
have very similar neutron cross sections. Monte Carlo simulations
show that CF$_{3}$I and C$_{3}$F$_{8}$ or C$_{4}$F$_{10}$ exhibit
essentially the same response to any residual neutron background,
i.e., neutrons cannot mimic an observed behavior such as that
described in the discussion of Fig.\,\ref{targets}. Other
combinations of targets such as germanium and silicon are
more prone to systematic effects where residual neutron recoils can
mimic the response expected from a WIMP with dominant 
spin-independent couplings.

The arguments presented in Ref.~\cite{Bertone:2007xj} can be easily generalized to a 
combination of data from
experiments using targets maximally sensitive to different
couplings, supporting the tenet that a large variety of dark matter
detection methods is presently desirable.

\section{Astrophysical data: from hints to smoking-guns}

The difficulty of obtaining from astrophysical
observations conclusive answers on the nature of DM, is 
witnessed by the numerous conflicting claims of 
discovery, recently appeared in literature. A number of 
observations have been in fact ``interpreted'' in terms of
DM, without providing, though, conclusive enough evidence
to claim ``discovery''. The reason is simple: our understanding 
of the nature and distribution of DM is so poor that 
we have enough freedom in the choice of physical parameters,
such as the DM mass and annihilation rate (roughly speaking
equal to the product of the annihilation cross-section
times the integral of the density squared), to fit any 
unexplained "bump" observed in astrophysical spectra. 

Examples of possible hints of discovery recently appeared 
in literature include (see also the discussion in Refs.~\cite{Bertone:2006nw,Hooper:2007vy}):

\begin{itemize}

\item{{\bf MeV Dark Matter.}} The INTEGRAL observation
of an intense 511 keV annihilation line from a region of
size $\approx 10 ^\circ$ centered around the galactic 
center~\cite{Knodlseder:2005yq} has reopened an old debate on
the origin of the population of positrons observed in the Galactic
bulge.

\begin{figure*}[t]
\centering
  \includegraphics[width=0.45\textwidth]{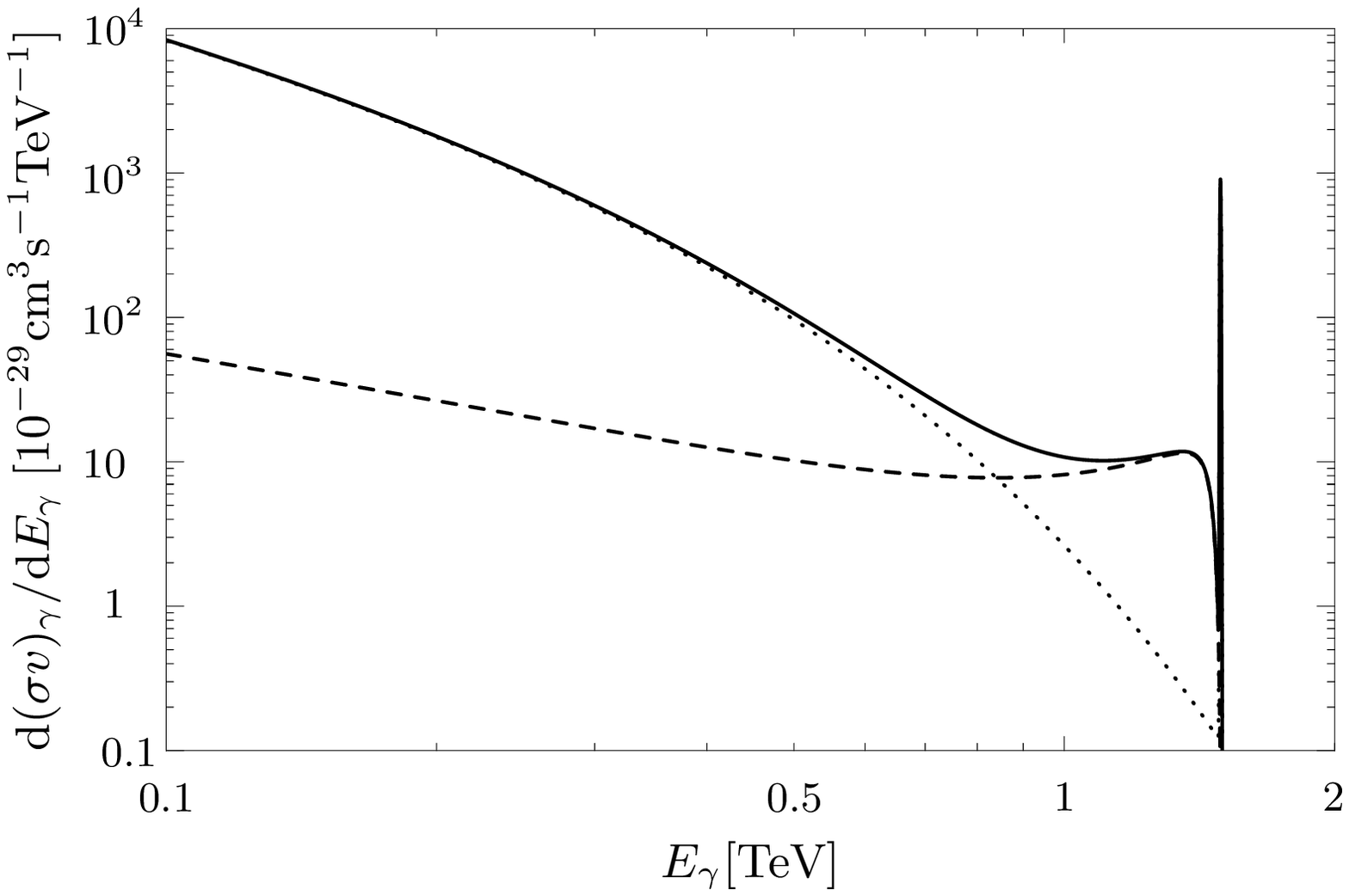}
  \includegraphics[width=0.45\textwidth]{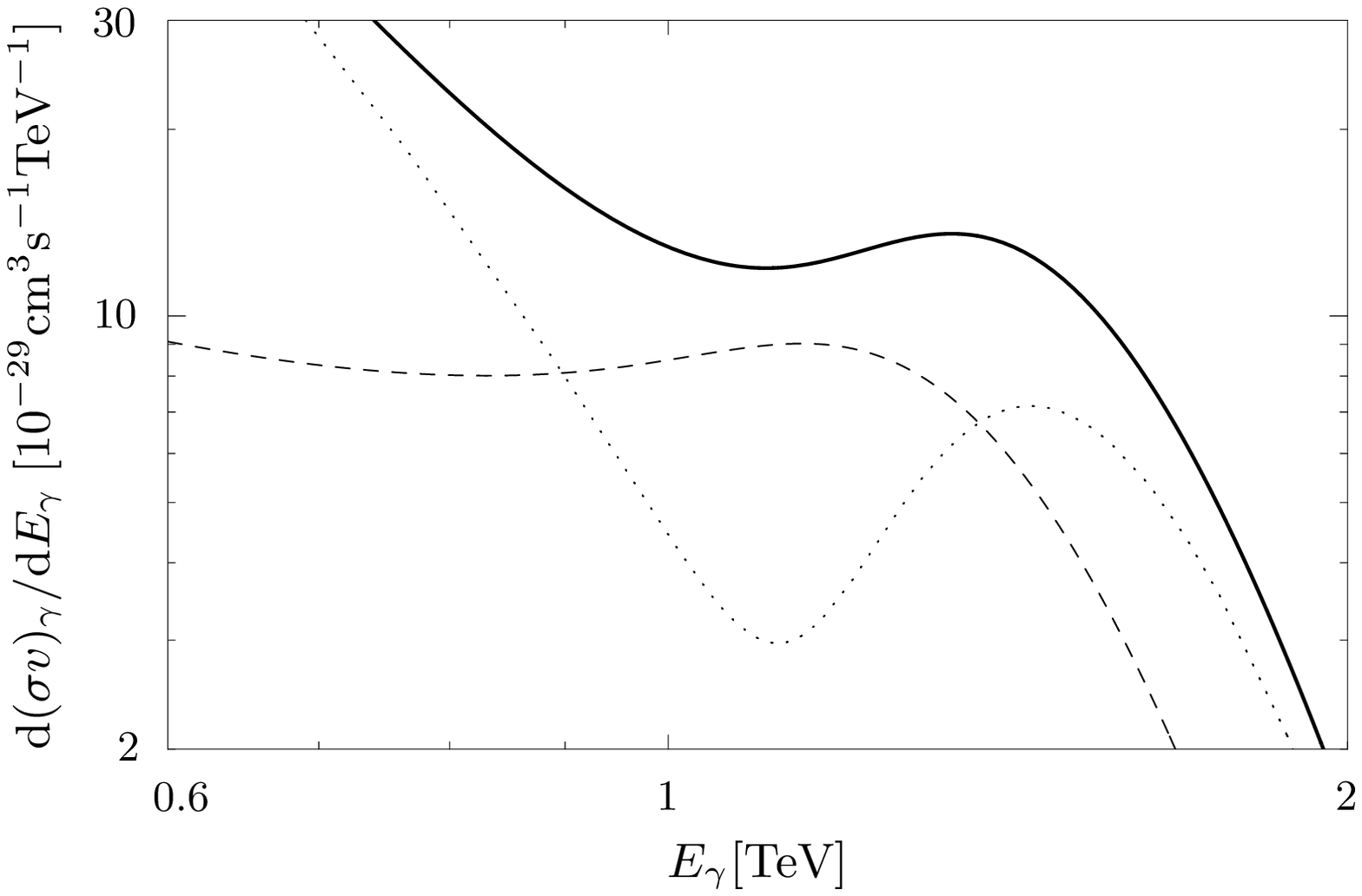}
\caption{{\it Left Panel:} The total differential photon distribution 
for a heavy neutralino.
Also shown separately is the contribution from radiative processes
$W^+W^-\gamma$ (dashed), and the $W$
fragmentation together with the $\gamma\gamma,
\, Z\gamma$ lines (dotted).{\it Right Panel:} Same spectra 
as seen by a detector with an energy resolution of 15 percent. Both figures from Ref.~\cite{Bergstrom:2006hk}}
\label{fig:spectra}       
\end{figure*}
The large uncertainties associated with the many astrophysical 
explanations proposed in the literature have left the door open to more
``exotic'' explanations. In particular, the possibility to explain the data 
in terms of DM annihilations immediately 
attracted the attention of particle astrophysicists. 
A simple calculation, however, suggested that
any candidate with a mass above the pion mass would inevitably 
produce gamma-rays and synchrotron emission far above the 
experimental data. In particular, if the 511 keV emission was
due to positrons produced by annihilation of neutralinos, 
the associated gamma-ray flux would exceed the observed EGRET 
flux by seven orders of magnitude. A Light DM candidate 
was instead shown to successfully reproduce the normalization
of the observed 511 keV line without violating any other 
observational constraint~\cite{Boehm:2003bt}. The 
Light DM interpretation is to be considered {\it tentative} until
one can find a smoking-gun for it, or make a testable prediction.
The first prediction, i.e. the detection of an annihilation line
from a dwarf galaxy, has so far failed~\cite{Cordier:2004hf}
, while further analyses
have progressively reduced the allowed parameter space of DM 
particles. On one side, an upper limit on the mass comes from
the analysis of Internal Bremsstrahlung emission ($\approx 20$ MeV, see 
Ref.~\cite{BBB05}) and in-flight annihilation (of order $3-7$ MeV, see Refs. 
~\cite{BY06,Sizun:2006uh}). 
On the other side, an analysis based on the explosion of the
supernova SN1987A sets a lower limit of $\approx 10$ MeV,
thus apparently ruling out Light DM as a viable explanation 
of the 511 keV line~\cite{Fayet:2006sa}, at least in its 
most simple realization.
Recently, these constraints have been 
challenged, and the claim has been made that it is still possible
to accommodate all existing constraints while still providing
a satisfactory explanation of the INTEGRAL data~\cite{Boehm:2006df}:
the debate is thus still open. Peculiar spectral 
features such as a 2$\gamma$ line~\cite{Boehm:2006gu}), or 
discovery in collider searches would allow to promote the 
Light DM scenario 
from ``tentative interpretation'' to ``discovery''.

\item{{\bf The GeV Excess}} Evidence for WIMPs
with a mass of tens of GeV, producing through their annihilation
a ``bump'' in the Galactic gamma-ray emission observed by EGRET 
was recently claimed in Ref. ~\cite{deBoer:2006tv}. 
Although in principle very exciting, the emission is characterized by
a distribution which is
very different from the one na\"ively predicted by numerical simulations
(more intense towards the galactic center), being in the shape
of a ring around the galactic center. This is not sufficient of course to rule 
out this scenario, but there are still numerous difficulties
associated with this intepretation, that have been recently 
highlighted in Ref.~\cite{Bergstrom:2006tk}, in particular regarding the required
ring-shaped distribution of DM, as well as the apparent 
incompatibility with anti-proton measurements.  
As in the case of MeV DM, this doesn't mean that
the proposed interpretation is wrong, but simply that a 
different approach is needed to obtain conclusive evidence.

\begin{figure*}[t]
\centering
  \includegraphics[width=0.4\textwidth]{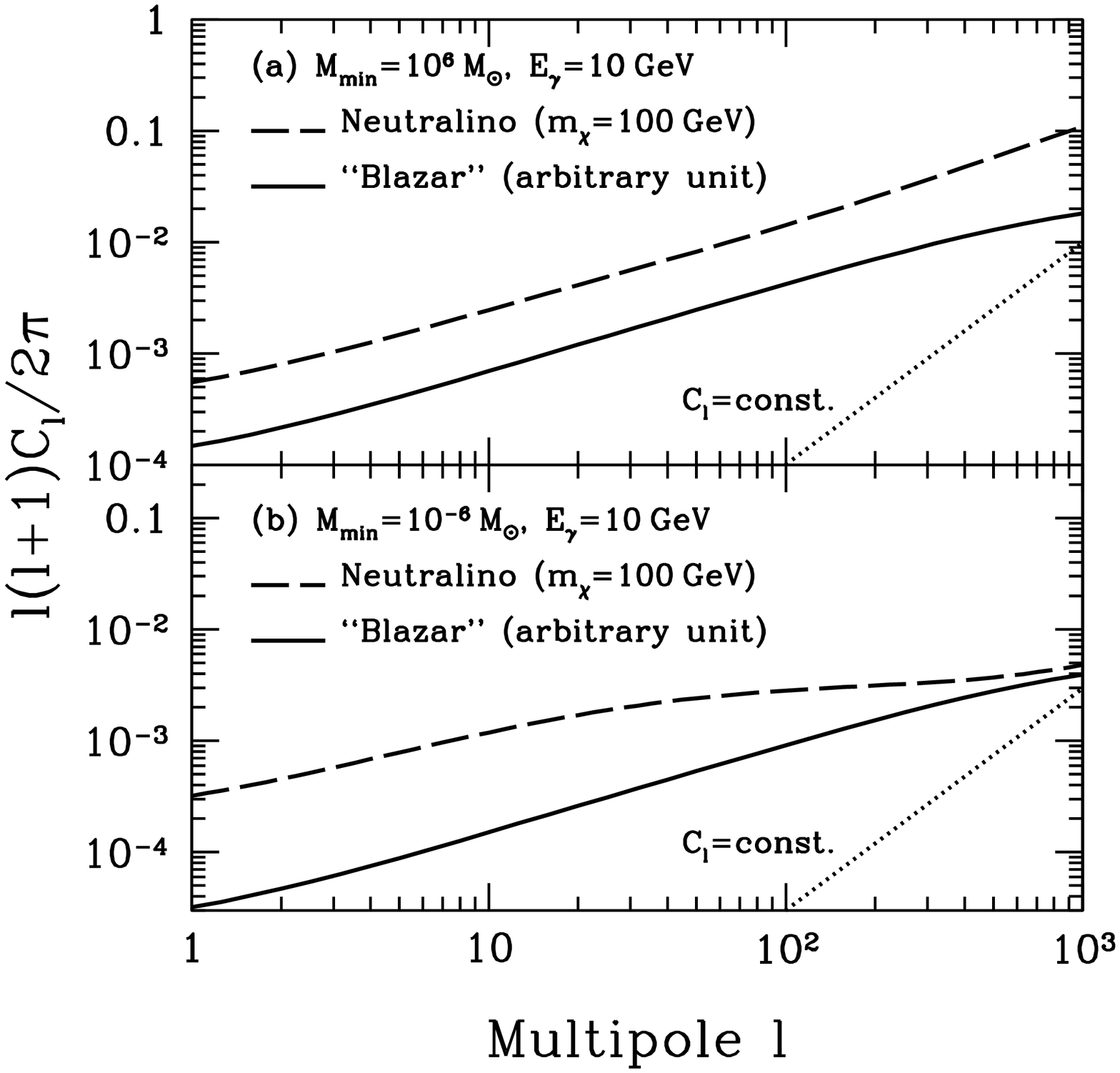}
  \includegraphics[width=0.45\textwidth]{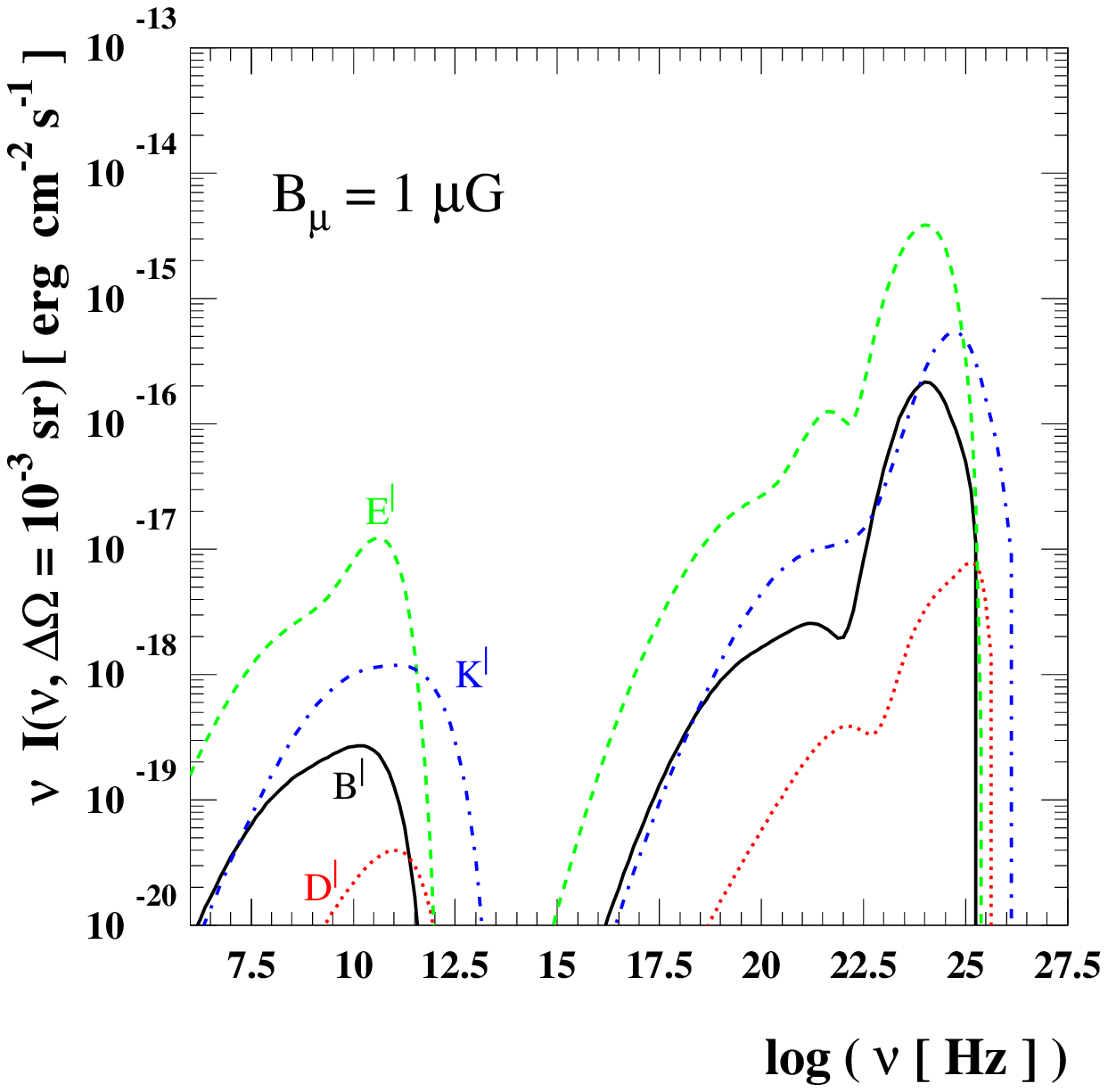}
\caption{{\it Left Panel:} Shape of the angular power spectrum of the CGB expected 
 from unresolved blazar-like sources
 (solid lines)
 with arbitrary normalizations.
 The power spectrum from annihilation of neutralinos 
 with $m_\chi = 100$ GeV is also plotted as the dashed lines. 
 The adopted gamma-ray energy is
 10 GeV, and the minimum mass of dark matter halo 
 is (a) $10^6 M_\odot$, and (b) $10^{-6} M_\odot$. 
 The dotted lines show the shot 
 noise ($C_l={\rm const.}$) with arbitrary  normalizations,
 which represent the power spectrum of very rare sources.
 From Ref.~\cite{Ando:2005xg}. {\it Left Panel:}Multi-wavelength spectra for four different benchmark DM models,
for a best fit NFW profile, and a mean magnetic field equal to $1 \mu G$. 
From Ref.~\cite{cola1}, see {\it ibid.} for more details.}
\label{fig:combi}       
\end{figure*}
\item{{\bf The Galactic center.}} The discovery of a gamma-ray source
in the direction of Sgr A* has long been considered a potentially perfect 
signature of the existence of particle DM, as thoroughly discussed
in Refs.~\cite{Stecker:88,Bouquet:89,Berezinsky:94,Bergstrom:97,Bertone:2001jv,Cesarini:2003nr,Fornengo:2004kj}. However, the gamma-ray source observed by EGRET at the Galactic center
might be slightly offset
with respect to the position of Sgr A*, a circumstance clearly
at odds with a DM interpretation~\cite{Hooper:2002ru}.

Recently the gamma-ray telescope HESS has detected a high energy 
source, spatially coincident within $1'$ with Sgr A*~\cite{Aharonian:2004wa}
and with a spectrum extending above 20 TeV. 
Although the spatial coincidence is much more satisfactory 
than in the case of the EGRET source, the ``exotic'' origin
of the signal is hard to defend, since the implied mass scale 
of the DM particle (well above 20 TeV, to be consistent with the
observed spectrum) appears to be difficult to reconcile with 
the properties of commonly studied candidates , and the fact that 
the spectrum is a power-law, then, points towards a standard
astrophysical source (see e.g. the discussion
Ref.~\cite{Profumo:2005xd}). The galactic center,
however, remains an interesting target for GLAST, since it 
will explore a range of energies below the relatively high 
threshold of HESS, where a DM signal could be 
hiding~\cite{Zaharijas:2006qb}. The recent claim that the profile of 
large galaxies could be much more shallow than previously 
thought~\cite{Mashchenko:2006dm}, should not discourage
further studies, especially in view of the possible enhancement
of the DM density due to interactions with the stellar 
cusp observed at the Galactic center~\cite{MHB}. 

\item {\bf WMAP Haze.}
Aside from the expected astrophysical foregrounds, 
the WMAP observations have revealed an excess of microwave emission
in the inner $20^\circ$ around the center of the Milky Way.
This origin of this WMAP "Haze" is unknown, and 
conventional astrophysical explanations such as 
thermal Bremsstrahlung from hot gas, thermal or 
spinning dust, and Galactic synchrotron appear to 
be unlikely. Dark matter annihilations have been 
suggested as a possible explanation~\cite{haze1,haze2}, and if this is
the case, GLAST may find an associated gamma-ray signal ~\cite{haze3}.

\end{itemize}

Despite the difficulties associated with most of these strategies, 
and despite the lack of conclusive evidence, 
all these claims should be taken seriously and further investigated 
without prejudice, especially in view of the fact that we don't know what 
DM is. At the same time, it is important to look for clear smoking-gun 
of DM annihilation, and study theoretical scenarios with 
unambiguous signatures that can be tested 
with present and future experiments. To this aim, we summarize here
some recently proposed ideas that go precisely in this direction,
and that may shed new light on the nature of particle DM.

\begin{figure*}[t]
\includegraphics[width=0.9\textwidth]{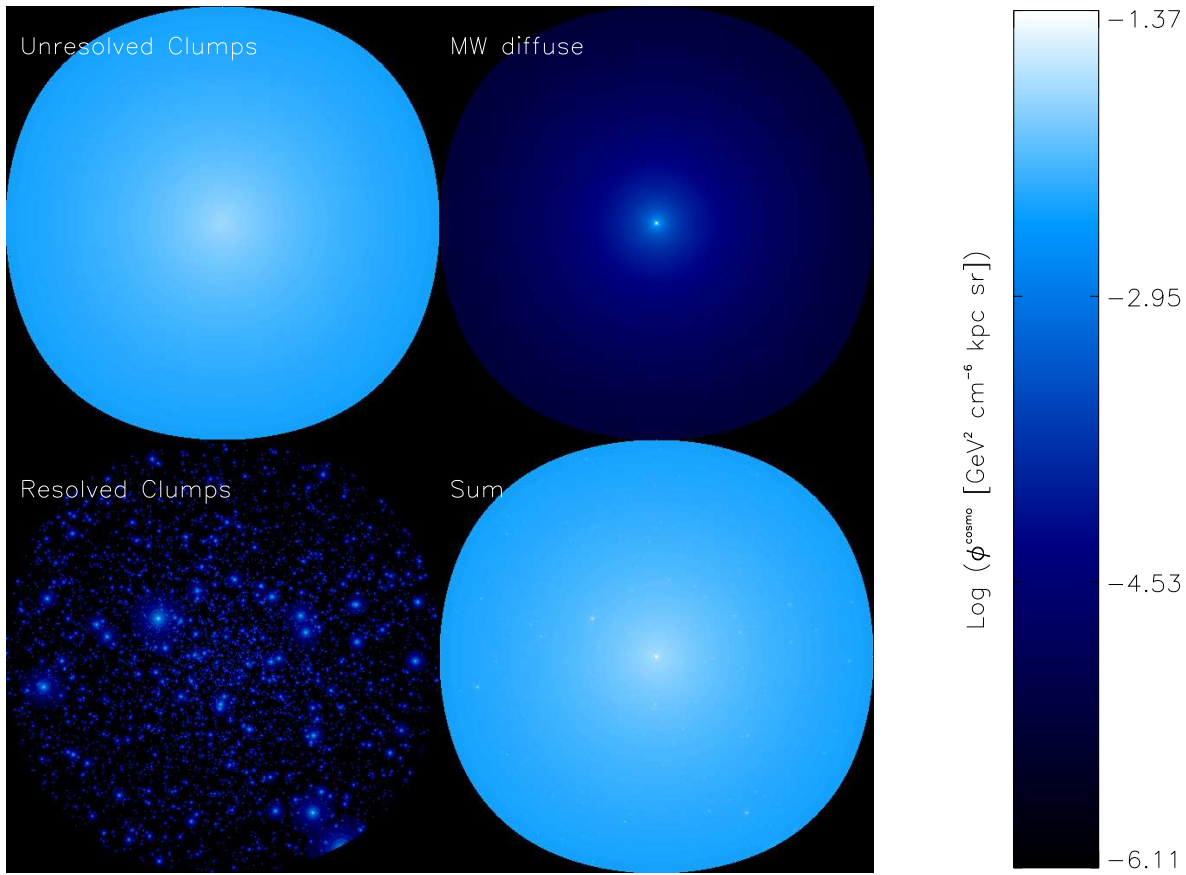}
\caption{Map of $\Phi^{\rm cosmo}$ (proportional to the annihilation signal) for the Bullock et al. model, in a cone of $50^\circ$ around the Galactic Center, as seen from the position of the Sun. Upper left: smooth subhalo contribution from unresolved halos. Upper right: MW smooth contribution. Lower left: contribution from resolved halos. Lower right: sum of the three contributions. From Ref.~\cite{Pieri:2007ir}.
}
\label{fig:clumps}
\end{figure*}

\paragraph{{\bf Spectral Features.}}
The first, and  more
clear signature that one may hope to detect is to identify 
distinctive spectral features in the DM annihilation spectra. 

It might be possible for instance to detect annihilation lines at an
energy equal to the DM particle mass. In the case of SUSY, although there are no tree level processes for neutralino annihilation into photons, loop level processes to $\gamma \gamma$ and $\gamma Z^0$ are very interesting, and may provide a spectral line feature observable in indirect detection experiments~\cite{bergstromgammagamma,bergstromgammaz}. Similar calculations have been performed for other candidates such as the aforementioned $B^{(1)}$ in theories with Universal Extra Dimensions~\cite{Bergstrom:2006hk}, and the so-called Inert Higgs DM \cite{Gustafsson:2007pc}.

Aside from the rather featureless gamma-ray spectrum produced by the fragmentation
of gauge bosons and quarks, there are additional, more distinctive, sources of photons. Internal Bremsstrahlung of W pair final states at energies near the mass of the neutralino appears particularly promising. For masses larger than about 1 TeV it results in a characteristic signal that may dominate not only over the continuous spectrum from W fragmentation, but also over the $\gamma-\gamma$ and $\gamma-Z$ line signals~\cite{Bergstrom:2005ss}. In figure 
~\ref{fig:spectra} we show the importance of radiative corrections $W^+W^-\gamma$ for the case of a heavy neutralino, as compared with line signals. Also shown are the same spectra convolved with the energy resolution of a GLAST-like detector.

Recently,  Bringmann et al. ~\cite{Bringmann:2007nk} have computed electromagnetic radiative corrections to all leading annihilation processes in the MSSM and mSUGRA, and pointed out that in regions of parameter space where there is a near degeneracy between the dark matter neutralino and the tau sleptons, radiative corrections may boost the gamma-ray yield by up to three or four orders of magnitude. This turns out to be true even for neutralino masses considerably below the TeV scale, and leads to a sharp step at an energy equal to the mass of the dark matter particle. For a considerable part of the parameter space, internal Bremsstrahlung appears then more important for indirect dark matter searches than line signals~\cite{Bringmann:2007nk}.

The possibility to discriminate an annihilation signal from ordinary astrophysical
sources has been addressed in Refs.~\cite{Baltz:2006sv,Bertone:2006kr}. 
If DM annihilation signals are within the reach of GLAST, the observation
of the high-energy cut-off would allow a measurement of the DM particle mass 
with an accuracy equal to the energy resolution of the experiment, i.e. $\Delta E / E \approx 10$\%~\cite{Bertone:2006kr}.

\paragraph{{\bf Gamma-ray background.}}
The first calculation of the gamma-ray background produced
by the annihilations of DM in all structures, at any redshift, was performed 
in Ref.~\cite{Bergstrom:2001jj} , and then further studied in Refs.~\cite{Taylor:2002zd,Ullio:2002pj}. 
The annihilation background can be expressed as
\begin{equation}
\Phi(E)= \frac{\Omega_{DM}^2 \rho_c^2}{8 \pi H_0}  \frac{\sigma v}{m_\chi^2}
\int_0^{z_{max}} \mbox{d}z \frac{\Delta^2}{h(z)} N(E')
\end{equation}
where $N(E')$ is the gamma-ray spectrum per annihilation,
$H_0$ is the Hubble parameter, $E'=E(1+z)$ and $h(z)=[(1+z)^3\Omega_{DM}+\Omega_\Lambda]^{1/2}$.
The information on the shape of individual DM halos in 
encoded in $\Delta^2$, which is essentially the integral 
of $\rho^2$ over the virial volume of the halo. 
Although it is unlikely that the annihilation
background will be detected without first detecting a prominent 
gamma-ray source at the Galactic center~\cite{Ando:2005hr}, the
characteristic power spectrum of the gamma-ray background
would discriminate its DM origin from ordinary astrophysical
sources~\cite{Ando:2005xg}.

We show in the left panel of fig.~\ref{fig:combi}   
 the power spectrum of the gamma-ray background produced 
by annihilation of neutralinos 
 with $m_\chi = 100$ GeV, compared with the one relative to 
 unresolved blazar-like sources.
Above $l\sim 200$ the DM spectrum continues to grow whereas 
the blazar spectrum flattens out,
due to the cut-off adopted by the authors corresponding to
 the minimum mass of halos hosting blazars ($\approx  10^{11} M_\odot$).
The annihilation spectrum thus appear to have much more power
at large angular scales, which should be easily distinguished
from the blazar spectrum. 

There are large uncertainties associated with
this calculation, mainly due to our ignorance of the DM profile 
in the innermost regions of halos, and of the amount of substructures.
The existence of mini-spikes (see below) would also dramatically affect 
the predicted result~\cite{Horiuchi:2006de}. But the clear prediction is
made that if the observed background has the peculiar shape discussed
above, this may be consider as a hint of DM annihilations.
Recently the calculation of the {\it neutrino} background from
DM annihilations has been performed, adopting a formalism very similar
to the one sketched above. The comparison with observational data 
allows to set an interesting, and very general, 
upper bound on the dark matter total annihilation cross section~\cite{Beacom:2006tt}.

\paragraph{{\bf Multi-messenger approach.}}
An alternative strategy is to employ a multi-messenger, multi-wavelength
approach. In fact, despite the freedom in the choice of DM
parameters makes the interpretation of observational data rather
inconclusive, one can always combine the information at different
wavelengths, and with different messengers, to obtain more
stringent constraints. In fact, gamma-rays are typically (but not
exclusively) produced
through annihilation and decay chain involving neutral pions
\begin{equation}
\chi \overline{\chi} \rightarrow q \overline{q} \rightarrow \left[ \mbox{fragmentation} \right] \rightarrow \pi^0 \rightarrow 2\gamma
\end{equation}
Every time gamma-rays are produced this way, leptons 
and neutrinos are also produced following the chain
\begin{equation}
\chi \overline{\chi} \rightarrow q \overline{q} \rightarrow \left[ \mbox{fragmentation} \right] \rightarrow \pi^\pm \rightarrow l, \nu_l, ... 
\end{equation}
An example of this approach is the combined
study of the gamma-ray emission from the Galactic center and 
the associated synchrotron emission produced 
by the propagation of electron-positron pairs in the Galactic
magnetic field~\cite{Gondolo:2000pn,Bertone:2001jv,Bertone:2002ms,Aloisio:2004hy}. Similarly one can investigate what the flux of 
neutrinos would be, once the gamma-ray flux has been normalized to
the EGRET data~\cite{Bertone:2004ag}. 

One can also ask what the fate 
of the electron-positron pairs produced by DM annihilation is in 
dwarf galaxies and clusters of galaxies. An example of this 
approach can be found in Refs.~\cite{cola1,cola2}, where the authors study
the synchrotron and gamma-ray emission from Draco and from 
the Coma cluster. In the right panel of fig.~\ref{fig:combi} we show 
the multi-wavelength spectra of Draco, relative to four different 
DM benchmark models, assuming a NFW profile and a mean
magnetic field of $1 \mu G$.

\begin{figure*}[ht]
\centering
  \includegraphics[width=0.5\textwidth]{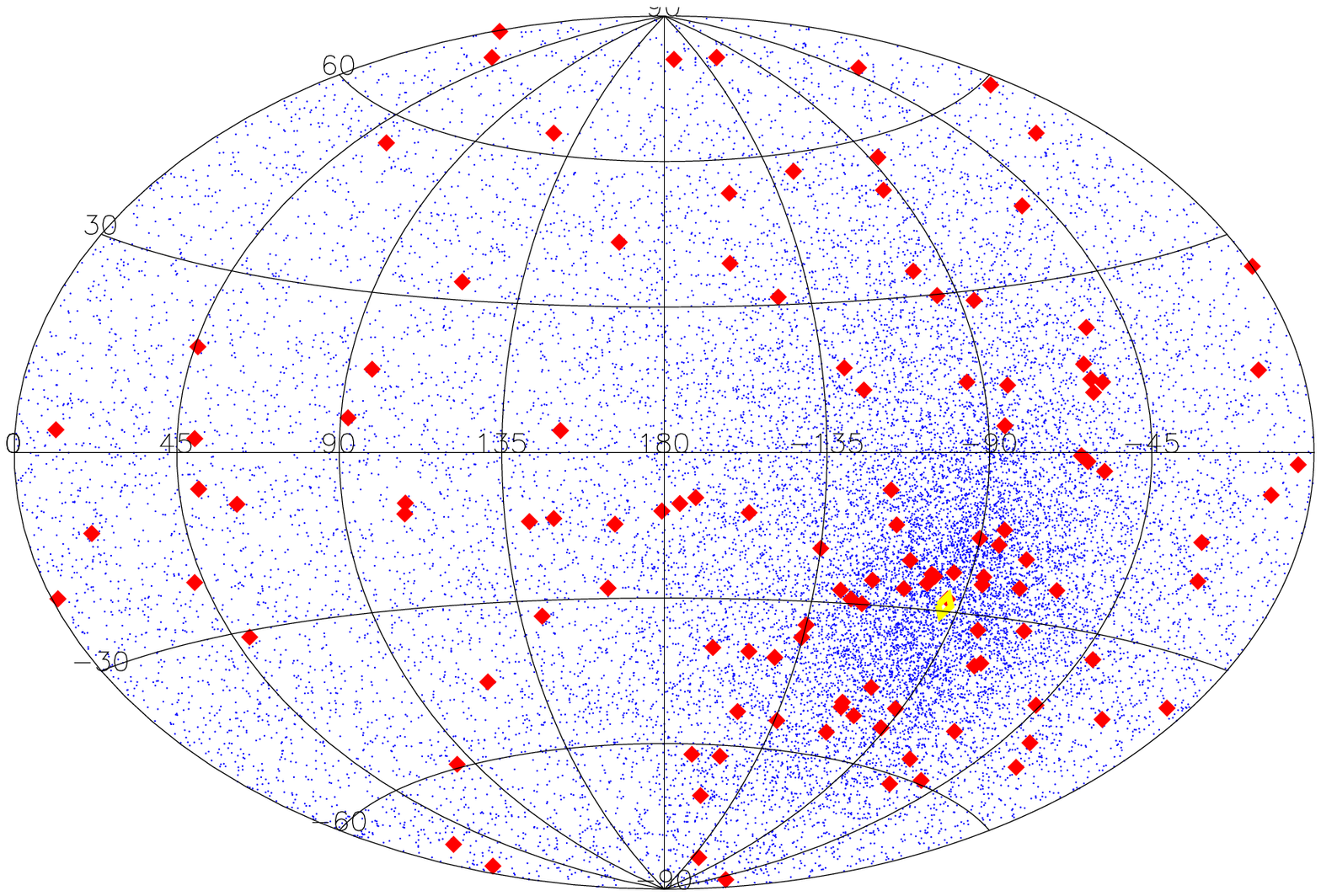}
  \includegraphics[width=0.3\textwidth]{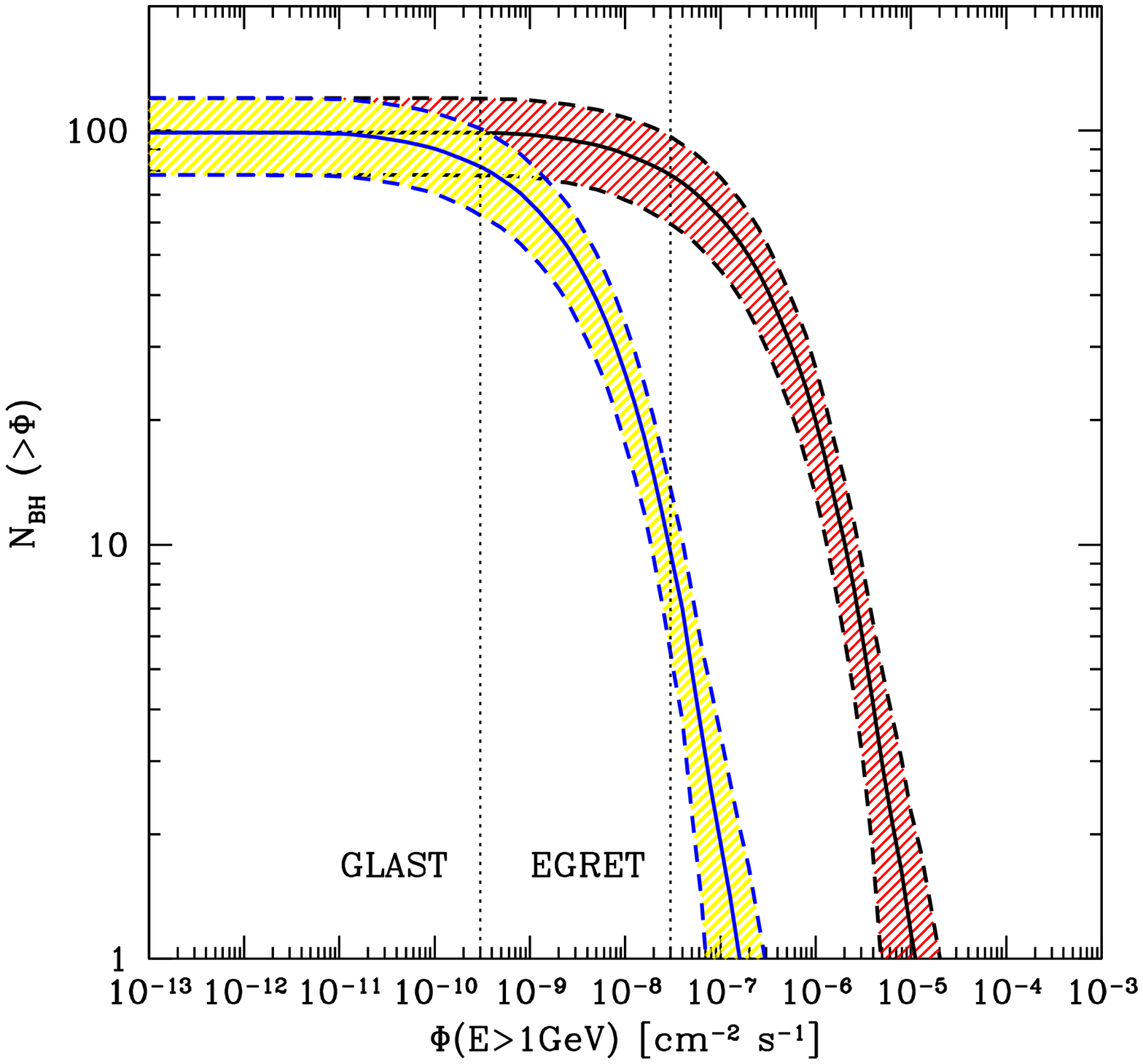}
\caption{{\it Left Panel:} Sky map in equatorial coordinates showing the position 
of Intermediate Mass Black Holes in one random realization
of a Milky-Way like halo (red diamonds), and in all 200 realizations (blue dots).
The concentration at negative declinations corresponds to 
the position of the Galactic center (black open diamond).
From Ref.~\cite{Bertone:2006nq}. {\it Right Panel:} IMBHs integrated luminosity function,
(number of mini-spikes detectable with an experiment of sensitivity
$\phi$) for IMBHs with mass $\sim 10^5 M_\odot$.
The upper (lower) line corresponds to $m_\chi=100$ GeV, $\sigma v=3\times 10
^{-26}$ cm$^3$ s$^{-1}$ ($m_\chi=1$ TeV, $\sigma v= 10
^{-29}$ cm$^3$ s$^{-1}$). For each curve we also show the 1-$\sigma$
scatter among different realizations of Milky Way-sized host DM halos.  
We show for comparison the 5$\sigma$ point 
source sensitivity above $1$~GeV of EGRET and GLAST (1 year). From Ref.~\cite{Bertone:2005xz}.}
\label{fig:last}       
\end{figure*}

A word of caution is in order, however, when combining information
relative to different wavelengths. In fact, not only the available
data, due to the different angular resolution of experiments, are
relative to different physical regions, but the calculation of
the associated spectra at different energies usually requires 
further inputs, thus introducing new parameters to the problem.
The aforementioned calculation of the synchrotron emission is a 
typical example: although for every specific DM model the
 number of electron-positron pairs produced per annihilation is
fixed, the calculation of the synchrotron emission requires an estimate 
of the diffusion of positrons and it further depends on the magnetic
field profile, typically poorly constrained on the scales of interest.

\paragraph{{\bf Clumps.}}
The detectability of individual DM substructures, or
”clumps”, has been widely discussed in literature, often with 
contradictory results. The number
of detectable clumps with a GLAST-like experiment, at
5$\sigma$ in 1 year and for a WIMP DM particle in fact
ranges from $\lesssim 1$~\cite{Koushiappas:2003bn} to more
than 50~\cite{Baltz:2006sv} for large mass halos, while for microhalos
(i.e. clumps with a mass as small as $10^{-6}$M$_\odot$) the predictions range from no detectable objects~\cite{Pieri:2005pg}
to a large number of detectable objects, with a fraction of
them exhibiting a large proper motion~\cite{Koushiappas:2006qq}.
The apparent inconsistency of the results published so far,
is actually due to the different assumptions that different
groups adopt for the physical quantities that regulate the
number and the annihilation ”brightness” of DM clumps.

In particular, even in the context of the benchmark density
profile introduced by Navarro, Frenk and White 1996
[NFW], the results crucially depend on the substructures
mass function, their distribution within the halo host and
their virial concentration $c(M, z)$ which is a function of mass
and of collapse redshift of DM clumps.

It was recently shown in Ref.~\cite{Pieri:2007ir}, that scenarios leading to a high
number of detectable sources, as well as scenarios where micro-clumps (i.e. clumps
with mass as small as $10^{-6}$M$_\odot$) can be detected, are severely constrained by the diffuse gamma-ray background detected by EGRET. For a fiducial DM candidate with mass
$m_\chi=100$ GeV and annihilation cross section $\sigma v = 10^{-26}$ cm$^3$ s$^{-1}$, 
at most a handful
of large mass substructures, and no micro-clumps, can be detected (at 5$\sigma$, with a
1-year exposure time, by a GLAST-like experiment) in the most optimistic scenario.
We show in fig.~\ref{fig:clumps} a map of a quantity defined in Ref.~\cite{Pieri:2007ir}, which is proportional to the annihilation signal for a specific case (Bullock et al. concentration model), in a cone of $50^\circ$ around the Galactic Center, as seen from the position of the Sun. The three contributions discussed in the paper are shown: a smooth one from unresolved subhalos; the MW smooth halo; and resolved individual halos. The sum of the three contributions is shown in the lower right panel.

\paragraph{{\bf Mini-Spikes.}}
Black Holes (BHs) can be broadly divided in 3 different classes. The first
class include BHs with mass smaller than $\approx 100$
solar masses, typically remnants of the collapse of 
massive stars (recent simulations suggest that the upper 
limit on the mass of these objects is as low 
as $\approx 20 M_\odot$ ~\cite{Fryer:2001}).
There is robust evidence for the existence of these objects,
coming from the observation of binary objects with
compact objects whose mass exceeds the critical mass of 
Neutron Stars. For a review of the topic and the discussion of the
possible smoking-gun for Stellar Mass BHs see e.g.~\cite{Narayan:2003fy} 
and references therein.

The existence of Supermassive BHs (SMBHs) , lying at the center of
galaxies (including our own), is also well-established 
(see e.g. Ref.~\cite{Ferrarese:2005}),
and intriguing correlations are observed between the BHs mass
and the properties of their host galaxies and 
halos~\cite{kormendy:1995,Ferrarese:2000se,McLure:2001uf,Gebhardt:2000fk,Tremaine:2002js,Koushiappas:2003zn}. From a theoretical point of view, a population of massive seed 
black holes could help to explain the origin of SMBHs. 
In fact, observations of quasars at 
redshift $z\approx 6$ in the 
Sloan Digital survey ~\cite{Fan:2001ff,barth:2003,Willott:2003xf}
suggest that SMBHs were already in place 
when the Universe was only $\sim 1$ Gyr old, a circumstance 
that can be understood in terms of rapid growth starting 
from ``massive'' seeds (see e.g. Ref.~\cite{haiman:2001}).  

This leads us to the third category of BHs, characterized 
by their {\it intermediate} mass. In fact, scenarios 
that seek to explain the properties of the observed 
supermassive black holes population  result in the prediction
of a large population of wandering Intermediate Mass BHs (IMBHs). Here, following 
Ref.~\cite{Bertone:2005xz}, we consider two different formation
scenarios for IMBHs. In the first scenario,
IMBHs form in rare, 
overdense regions at high redshift, 
$z \sim 20$, as remnants of Population III stars, and have a 
characteristic mass-scale of a few $10^2 M_\odot$ 
\cite{Madau:2001} (a similar scenario
was investigated in Ref.~\cite{Zhao:2005zr,islamc:2004,islamb:2004}).  
In this scenario, these black holes serve as the 
seeds for the growth supermassive black 
holes found in galactic spheroids \cite{Ferrarese:2005}.  
In the second scenario, IMBHs form directly out of cold gas 
in early-forming halos 
and and are typified by a larger mass scale, 
of order $10^5 M_\odot$~\cite{Koushiappas:2003zn}. 
In the left panel of Fig.~\ref{fig:last} we show the distribution of IMBHs
in the latter scenario, as obtained in Ref.~\cite{Bertone:2006nq}.

The effect of the formation of a central 
object on the surrounding distribution of matter has been investigated 
in Refs.~\cite{peebles:1972,young:1980,Ipser:1987ru,Quinlan:1995} 
and for the first time in the framework of DM annihilations in 
Ref.~\cite{Gondolo:1999ef}. It was shown that  the 
{\it adiabatic} growth of a massive object at the center of a 
power-law distribution of DM with index $\gamma$, induces 
a redistribution of matter into a new power-law (dubbed ``spike'') with index
$\gamma_{sp}=(9-2\gamma)/(4-\gamma)$
This formula is valid over a region of size $R_s \approx 0.2 r_{BH}$, 
where $r_{BH}$ is the radius of gravitational influence
of the black hole, defined implicitly as $M(<r_{BH})=M_{BH}$,
with $M(<r)$ mass of the DM  distribution within a 
sphere of radius $r$, and $M_{BH}$ mass of the Black Hole
~\cite{Merritt:2003qc}.
The process adiabatic growth is in particular valid for the SMBH at the 
Galactic center. A critical assessment of the formation {\it and survival}
of the central spike, over cosmological timescales, is presented
in Refs.~\cite{Bertone:2005hw,Bertone:2005xv} (see also references therein).
The impact of the spike growth and subsequent destruction on the
gamma-ray background produced by DM annihilations has been studied in Ref.
~\cite{Sein}.

Here we will not further discuss the spike at the Galactic center, 
and will rather focus our attention on {\it mini-spikes} around IMBHs.
If $N_{\gamma}(E)$ is 
the spectrum of gamma-rays per 
annihilation, the gamma-ray flux from an individual mini-spike 
can be expressed as~\cite{Bertone:2005xz}
\begin{equation}
\Phi_{\gamma}(E)  = \phi_0  m_{\chi,100}^{-2} (\sigma v)_{26} D_{\rm  kpc}^{-2} L_{\rm sp} N_{\gamma}(E)
\label{eq:intrinsic}
\end{equation}
with $\phi_0 = 9 \times 10^{-10} {\rm cm}^{-2}{\rm s}^{-1}$.  
The first two factors depend on the particle physics parameters,
viz. the mass of the DM particle in units of 100 GeV 
$m_{\chi,100}$, and its annihilation
cross section in units of $10^{-26} {\rm cm}^3/{\rm s}$, $(\sigma v)_{26}$,
while the third factor accounts for the flux dilution with 
the square of the IMBH distance to the Earth in kpc, $D_{\rm  kpc}$. 
Finally, the normalization of the flux is fixed by an adimensional 
{\it luminosity factor} $L_{\rm sp}$, that depends on the specific
properties of individual spikes. In the case where the DM profile 
{\it before} the formation of the IMBH follows the commonly
adopted Navarro, Frenk and White profile~\cite{Navarro:1996he},
the final DM density $\rho(r)$ around the IMBH will be described by a 
power law $r^{-7/3}$ in a region of size $R_s$
around the IMBHs. Annihilations themselves will set an upper limit
to the DM density $\rho_{\rm max}\approx m_\chi/[(\sigma v) t]$, where 
$t$ is the time elapsed since the formation of the mini-spike, and
we denote with $R_{\rm c}$ the ``cut'' radius where $\rho(R_{\rm c})=\rho_{\rm max}$. 
With these definitions, the intrinsic luminosity factor  
in Eq.~\ref{eq:intrinsic} reads
\begin{equation}
L_{\rm sp}\equiv
\rho^{2}_{100}(R_{{\rm s}}) R_{\rm s,pc}^{14/3}
R^{-5/3}_{\rm c,mpc}
\end{equation}
where $R_{{\rm s,pc}}$ and $R_{\rm c,mpc}$ denote respectively
 $R_{{\rm s}}$ in parsecs and $R_{\rm c}$ in units of $10^{-3}$pc,
$\rho_{100}(r)$ is the density in units of $100$GeV cm$^{-3}$.
Typical values of $L_{\rm sp}$ lie in the range 0.1 -- 10~\cite{Bertone:2005xz}. 

In the left panel of Fig.~\ref{fig:last}, we show the (average) integrated luminosity 
function of IMBHs in scenario B.  We define the integrated luminosity 
function as the number of black holes producing a gamma-ray 
flux larger than $\Phi$, as a function of $\Phi$. Loosely
speaking, this can be understood as he number of mini-spikes
that can be detected with an experiment with point source sensitivity $\Phi$ above 1~GeV.   
The upper (lower) line corresponds to $m_\chi=100$~GeV, 
$\sigma v=3\times 10^{-26}$~cm$^3$s$^{-1}$ 
( $m_\chi=1$~TeV, $\sigma v=10^{-29}$~cm$^3$s$^{-1}$).  
We show for comparison the point source sensitivity 
above 1~GeV for EGRET and GLAST, corresponding roughly 
to the flux for a $5\sigma$ detection of a high-latitude 
point-source in an observation time of 
1~year~\cite{Morselli:2002nw}.  
The dashed region
corresponds to the $1\sigma$ scatter between different 
realizations of Milky Way-sized halos.
This band 
includes the variation in spatial distributions of IMBHs 
from one halo to the next as well as the variation in 
the individual properties of each IMBH in each realization.

The implications of the mini-spikes scenario have been 
investigated by several authors. Their impact on
the gamma-ray background has been studied in Ref.~\cite{Horiuchi:2006de}, 
while the implications for anti-matter fluxes have been
derived in Ref.~\cite{Brun:2007tn}. A population of IMBHs similar to the one
derived for the Milky Way should also be present in other 
spiral galaxies similar to our own; the prospects for 
detecting IMBHs in M31, i.e. the Andromeda Galaxy, have been
studied in ~\cite{Fornasa:2007ap}. 

\begin{acknowledgments}

It is a pleasure to thank the organizers of Lepton Photon 2007, including the
young volunteers, for the excellent organization of the 
conference and the warm hospitality in Daegu, Korea.
We also thank Laura Baudis for her help in putting the various 
direct detection experiments on the world map shown in fig. 1.
This work is partially based on work done in collaboration
with John Beacom, Enzo Branchini, David Cerdeno, Juan Collar, Dan Hooper, 
Gregory Mack, David Merritt, Brian Odom, Lidia Pieri, Joe Silk and Andrew Zentner.

\end{acknowledgments}


\end{document}